**Modelling the Frequency of Home Deliveries: An Induced Travel Demand Contribution of Aggrandized E-shopping in Toronto during COVID-19 Pandemics**

**Yicong Liu,** M.A.Sc. Candidate
Department of Civil & Mineral Engineering
University of Toronto, Toronto, ON, Canada
Email: nora.liu@mail.utoronto.ca

**Kaili Wang,** M.A.Sc. Candidate
Department of Civil & Mineral Engineering
University of Toronto, Toronto, ON, Canada
Email: jackkaili.wang@mail.utoronto.ca

**Patrick Loa,** Ph.D. Student
Department of Civil & Mineral Engineering
University of Toronto, Toronto, ON, Canada
Email: patrick.loa@mail.utoronto.ca

**Khandker Nurul Habib,** Ph.D., P.Eng.
Percy Edward Hart Professor
Department of Civil & Mineral Engineering
University of Toronto, Toronto, ON, Canada
Email: khandker.nurulhabib@utoronto.ca

Word Count: 6980 words + 2 tables (250 words per table) = 7480 words

The paper was presented at 2022 Annual Meeting of Transportation Research Board

**ABSTRACT**

The COVID-19 pandemic dramatically catalyzed the proliferation of e-shopping. The dramatic growth of e-shopping will undoubtedly cause significant impacts on travel demand. As a result, transportation modeller's ability to model e-shopping demand is becoming increasingly important. This study developed models to predict households' weekly home delivery frequencies. We used both classical econometric and machine learning techniques to obtain the best model. It is found that socioeconomic factors such as having an online grocery membership, household members' average age, the percentage of male household members, the number of workers in the household and various land-use factors influence home delivery demand. This study also compared the interpretations and performances of the machine learning models and the classical econometric model. Agreement is found in the variable's effects identified through the machine learning and econometric models. However, with similar recall accuracy, the ordered probit model, a classical econometric model, can accurately predict the aggregate distribution of household delivery demand. In contrast, both machine learning models failed to match the observed distribution.

**1.0 Introduction**

The COVID-19 pandemic heightened the popularity of e-shopping everywhere. Because of the implementation of travel restrictions and the temporary closure of non-essential stores during the pandemic, e-shopping has increased in prevalence. In December 2016, e-shopping only constituted 3.4% of the total retail sales in Canada (1); this number increased to 8.6% in December 2020 during the pandemic. At the same time, the share of e-shopping in the U.S. expanded to 14.0% in the fourth quarter of 2020, compared to 11.0% pre-pandemic (2). As a convenience to customers, e-shopping is usually coupled with home delivery services. The dramatic growth of e-shopping will undoubtedly cause significant impacts on travel demand.

Two main areas are influenced by e-shopping: passenger travel demand and freight travel demand (3). The potential substitution effect of e-shopping on in-store shopping trips can reduce the individual's demand for shopping trips. On the other hand, the increasing amount of delivery trips induced by e-shopping can alter the demand and travel patterns of freight trips. Understanding the changes in individual and freight travel behaviour caused by e-shopping demand can significantly impact travel demand and congestion management. Thus, the ability to model e-shopping demand becomes increasingly crucial for transportation modellers. Several studies have investigated the relationship between e-shopping and in-store shopping and their impacts on travel demand; however, very few have attempted to model and forecast e-shopping demand and home delivery frequencies.

It is important to find the best possible method to model the e-shopping and home delivery demand. However, such modelling is challenging without having clear information on the behaviour of the shippers and parcel delivery companies. Therefore, we examine modelling techniques from the fields of classical econometric modelling and machine learning (ML), given that ML models have received increasing attention and have been applied in various transportation studies (4-9). For classical econometric models, the ordered probit model is chosen because the measurement of delivery demand is ordered in nature in our dataset. For ML models, we use random forest (RF) and XGBoost, two contemporary and popular tree-based ensemble models. The performances of all models are compared based on prediction capacity. Moreover, to facilitate effective policymaking and planning efforts, correctly capturing causality and behavioural knowledge is equally essential as building models. Thus, the interpretation of variables and their impacts on model results are also compared and discussed across all three models. The factors that influence households' e-shopping and home delivery demand are investigated and discussed in the study. This paper fills the gap in literature on modelling home delivery demand, at the same time provides a systemic comparison between popular ML models and econometric models for modelling home delivery demand.

The remainder of the paper is organized as follows. Section 2 presents a literature review of the studies relevant to e-shopping demand and behaviour. Section 3 explains the methodology of the models used to predict home delivery frequencies. Section 4 describes the data used and summarizes descriptive statistics. Section 5 presents modelling results and post-modelling analysis on variable interpretation. Section 6 concludes the study.

## 2.0 Literature Review

Transportation researchers have noticed the rapid growth of e-shopping and started to study its impact on travel behaviours. Early research has been focused on investigating interaction patterns between e-shopping and in-store shopping activities. Commonly identified effects between e-shopping and in-store shopping are the substitution effect and complementarity effect. E-shopping is believed to potentially replace in-store shopping, thereby reducing shopping trips (3). Cao (3) provided a comprehensive summary of the early literature on the topic. These studies found that, although there was evidence of a substitution effect between online and in-store shopping, the magnitude of the effect was not statistically significant. Meanwhile, there were also studies that observed complementarity effects between online and in-store shopping (10,11).

It is worth noting that most of the earlier studies only considered durable goods (e.g., books, clothing, electronic devices). E-shopping studies regarding fast-moving consumer goods (FMCGs), namely groceries, was rarely studied until very recently. FMCGs are defined as inexpensive, frequently purchased goods such as perishable goods (e.g., meats, fruits), packed foods, and daily products (12). They are usually purchased more frequently than durable goods. This indicates shopping behaviour regarding FMCGs might have more significant impacts on travel demand than durable goods. There were several recent studies that examined shopping behaviour regarding FMCGs and groceries. Suel et al. (13) identified net substitution effects between online and in-store grocery shopping. The substitution pattern was later confirmed by Suel et al. (14) through a hazard-based model. They discovered that the probability of in-store grocery shopping could be dramatically decreased by engaging in online grocery shopping.

Some studies also explored the factors that might impact the frequency of online shopping and home deliveries. Cao (15) analyzed internet users' shopping behaviour on search goods (e.g., books, CDs, albums, etc.) and concluded that their choices of shopping channels are largely influenced by factors including product awareness, the convenience of searching for product-related information online, and product trial. Zhai et al. (16) also included experience goods (i.e., clothing) in their study to compare and found that consumers are more likely to choose in-store shopping for experience goods. Ramirez (17) also concluded that residents in urban areas generally performed more online shopping activities. He also discovered that millennials, women, individuals from higher-income households, those with greater levels of educational attainment, and those who tend to commute using active modes and public transit are more likely to shop online than in-store.

However, very few studies have attempted to develop models that predict the frequency of online shopping activities. Dias et al. (18) jointly modelled the weekly frequency of various in-store and online (with home delivery) shopping activities, distinguished by grocery, non-grocery, and cooked meals, using a multivariate ordered probit model. They found that socio-demographic variables can impact the frequency of online and in-store shopping activities, including income, employment status, type of housing unit, residential density, and vehicle ownership.

Barua et al. (19) adopted ML models to predict the frequency of online shopping activities at the household level. They concluded that household income, household size, neighbourhood density, and age and education level of household members are determinants of home delivery demand.

They utilized two tree-based ensemble models (RF and gradient boosting machine) and compared their predictive performance with linear and quadratic regression models. Their study claimed the superiority of ML models over classical statistical models in terms of performance. Their results showed that the gradient boosting machine model was the best-performing model (19). However, the validity of the comparison is questionable, as the statistical model compared in their study is rather simplistic. They compared ML models with linear and quadratic regression models, which are among the most basic statistical models. Unfortunately, their study did not compare the interpretation of the ML model results horizontally with econometric models. As discussed previously, deriving behavioural knowledge should be as important as establishing the model itself. This study aims to provide a fair comparison between ML models and classical econometric models. In our study, we compare the performance between RF, XGBoost and ordered probit models, which is a more robust demand generation model compared to linear regression models. In addition, a horizontal comparison between results interpreted from all models is also performed in this study.

Interpretation of ML models is receiving increasing attention in the transportation demand modelling community. Interpretation of the classical econometric model results is well supported by various economic and behavioural theories. However, the interpretation of post-modelling analyses of ML models could be less behaviourally sound. In the field of travel demand modelling, Wang et al. (8) made early attempts to link behavioural knowledge to deep neural networks (DNNs). They found that behavioural knowledge can effectively guide the architecture of DNNs. They also calculated pseudo-elasticities in ML models and found that pseudo-elasticity generated by DNN with alternative-specific utility functions produced similar outcomes as discrete choice models (DCMs). Zhao et al. (9) also compared behavioural outcomes between ML models and DCMs. They concluded that RF models agreed with DCMs regarding the importance of independent variables and the direction of independent variables' effects on dependent variables. However, they also found that RF models produced unreasonable elasticities.

Nonetheless, empirical studies aiming to gain behavioural knowledge from ML models are still limited. More studies should aim to investigate this issue both from a theoretical and empirical standpoint. Previous works primarily focus on comparing ML models and DCMs, especially logit kernel models that entirely or partially follow the independence of irrelevant alternatives (IIA) assumption (7-9,20). None of the literature compares the performance between ML models and non-linear econometric models. This paper contributes to the literature in this regard.

## 3.0 Methodology
Two types of modelling approaches are estimated and empirically evaluated. These are machine learning and classical econometrics.

## 3.1 Machine Learning Models

## Model Development
The model development process includes two stages: training and testing. To improve the model fit, the hyperparameters of the models are also tuned during the training stage. This study utilizes the Randomized Search technique to look for the most suitable combination of hyperparameters for the model. In the Randomized Search technique, a bounded domain for each hyperparameter is set as the input value. Then a set of random combinations of the hyperparameter values within the domain are tested to obtain the combination that can achieve the highest prediction accuracy. To make sure the model is not subject to overfitting, 10-fold cross-validation is applied when training the model. Finally, the selected model is applied to the testing dataset to test its prediction accuracy further, with the testing results reported in the form of a confusion matrix.

The confusion matrix presents the recall and precision of each outcome class. The equations used to calculate recall and precision are shown below:

$$Precision = \frac{observations\ that\ are\ correctly\ predicted\ as\ outcome\ N}{all\ observations\ that\ are\ predicted\ as\ outcome\ N} \tag{1}$$

$$Recall = \frac{observations\ that\ are\ correctly\ predicted\ as\ outcome\ N}{all\ observations\ from\ outcome\ N} \tag{2}$$

In this study, the prediction of household delivery frequency is treated as a classification problem. The ML models are to classify the household delivery frequency into six classes, each representing outcomes from zero to five or more deliveries.

**Random Forest Model**
Random forest (RF) is a supervised learning technique (21). A RF model is comprised of numerous independent decision trees. The decision trees are developed in parallel during the training process. A subset of the training data and the variables are randomly selected through bootstrap sampling to develop each decision tree. Each decision tree generates a prediction result that contributes to a majority voting system to determine the final result of the model. RF is designed to reduce variance in the model, and it is robust against overfitting.

**eXtreme Gradient Boosting (XGBoost) Model**
eXtreme Gradient Boosting (XGBoost) is an advanced implementation of the gradient boosting technique (22). It follows the basic element of a gradient boosting framework: to use an ensemble of shallow decision trees to minimize loss. In addition, it also utilizes regularization, parallel computing, and column sampling. Such additional techniques are incorporated to increase computational speed and reduce overfitting.

**3.2 Classical Econometric Model**

**Ordered Probit Model**
Ordered models are usually adopted when the dependent variable is categorized from ordinal measurements, and the categories may not be evenly spaced. It is a widely used model in travel demand modelling (20). In this study, the dependent variable is the weekly frequency of home delivery. The number of home deliveries per week are modelled as follows:

$$Y_i = \begin{cases} 0 \ if -\infty \leq Y_i^* \leq \tau_0 \ (no \ home \ delivery) \\ 1 \ if \ \tau_0 \leq Y_i^* \leq \tau_1 \ (1 \ home \ delivery) \\ 2 \ if \ \tau_1 \leq Y_i^* \leq \tau_2 \ (2 \ home \ deliveries) \\ 3 \ if \ \tau_2 \leq Y_i^* \leq \tau_3 \ (3 \ home \ deliveries) \\ 4 \ if \ \tau_3 \leq Y_i^* \leq \tau_4 \ (4 \ home \ deliveries) \\ 5 \ if \ \tau_4 \leq Y_i^* \leq +\infty \ (5 \ or \ more \ home \ deliveries) \end{cases} \quad (3)$$

where $\tau$ terms are the threshold parameters to be estimated by the model. The utility function ($Y_i^*$) is composed of a systematic utility ($\beta x_i$) and a random error term ($\varepsilon_i$) as shown below:

$$Y_i^* = \beta x_i + \varepsilon_i \quad (4)$$

where $x_i$ is the vector of observed variables, and $\beta$ is the parameter vector of $x_i$. The assumption for the error term of the ordered probit model is that it follows the standard normal distribution. The probability for the number of home deliveries per week is determined by the following equations:

$$\Pr(0 \ delivery) = \Pr(\varepsilon_i < \tau_0 - \beta x_i) \quad (5)$$

$$\Pr(n \ deliveries) = \Pr(\varepsilon_i < \tau_n - \beta x_i) - \Pr(\varepsilon_i < \tau_{n-1} - \beta x_i), \text{where n} \quad (6)$$
$$= 1,2,3,4$$

$$\Pr(5 \ or \ more \ deliveries) = \Pr(\varepsilon_i > \tau_4 - \beta x_i) \quad (7)$$

### 4.0 Data & Descriptive Statistics

The study used a subset of 851 records in the 2021 Summer iteration of the **COV**ID-19 influenced **H**ousehold's **I**nterrupted **T**ravel **S**chedules (COVHITS) survey. The COVHITS survey is an online household travel survey conducted in the Greater Toronto Area (GTA), Canada. The 2021 Summer COVHITS survey was conducted to understand people's travel behaviour and home delivery demand after the third wave of the COVID-19 pandemic in the study area. The subset of data was collected in July 2021. The survey randomly recruited samples from commercial survey panels managed by a market research company.

Distributions of key socioeconomic variables are compared to those from the 2016 Transportation Tomorrow Survey (TTS), a regional household travel survey that is conducted in the study area every five years (23). The distributions of key household attributes in the sample are shown in Figure 1. Overall, key household attributes in the dataset matched closely with the population characteristics in the study area. Figure 1 shows that distributions of essential household attributes all followed similar patterns as indicated in the 2016 TTS. However, there are a few discrepancies in household size distributions, household vehicle ownership, and household income between the two surveys. The sample used in this study contain a larger share of respondents from one- and two-person households than the 2016 TTS and a smaller percentage of respondents from households with four or more members. This can be attributed to the 2021 COVHITS survey using a commercial survey panel as its sampling frame, while 2016 TTS covered 5% of the entire population. Also, the distribution of household income is relatively similar between the two datasets, except for individuals from a household earning between $60,000 and $99,999 annually. Moreover, it is worth noting that 18% of households in 2016 TTS refused to report their income, whereas only 6% in the dataset declined to do so.

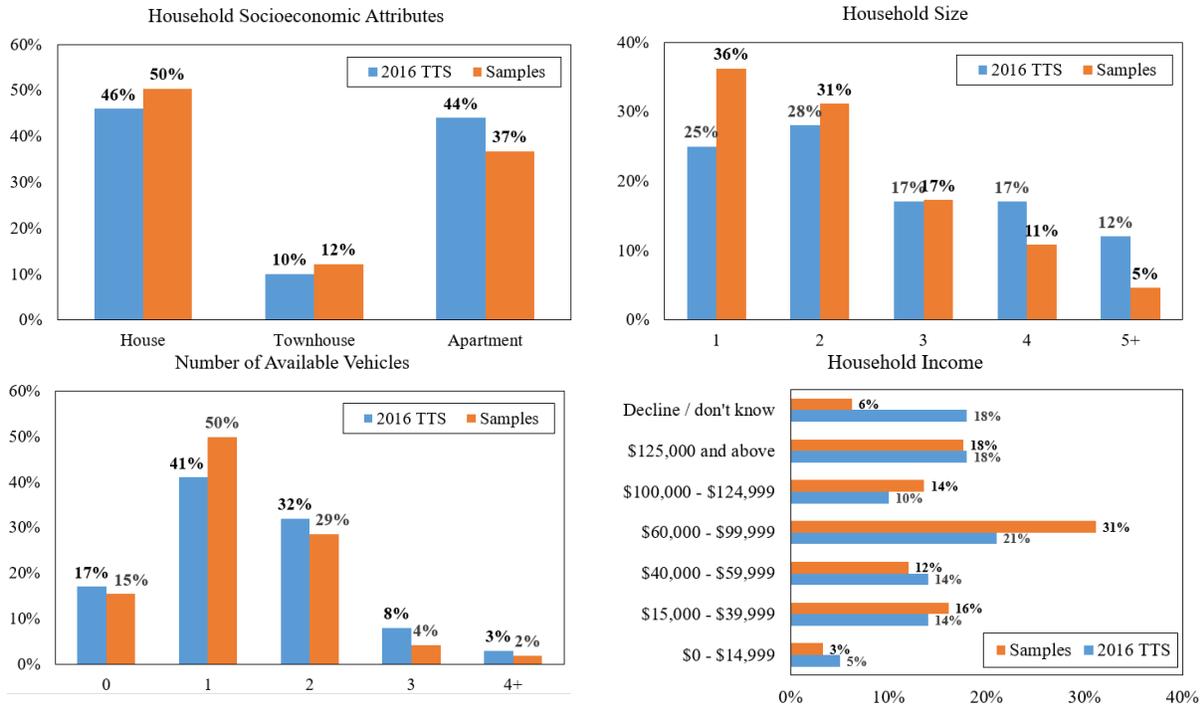

Figure 1. Comparison of key household attributes – sample vs. 2016 TTS

For personal attributes, a comparison of the distributions of the age and gender of the sample and the 2016 TTS are shown in Figure 2. The gender distribution in the sample is consistent with that of the 2016 TTS. In terms of age, the samples contain a slightly greater share of respondents aged between 18 and 64. However, the average age of the respondents (40 years old) is consistent with the average age of the 2016 TTS (38 years old).

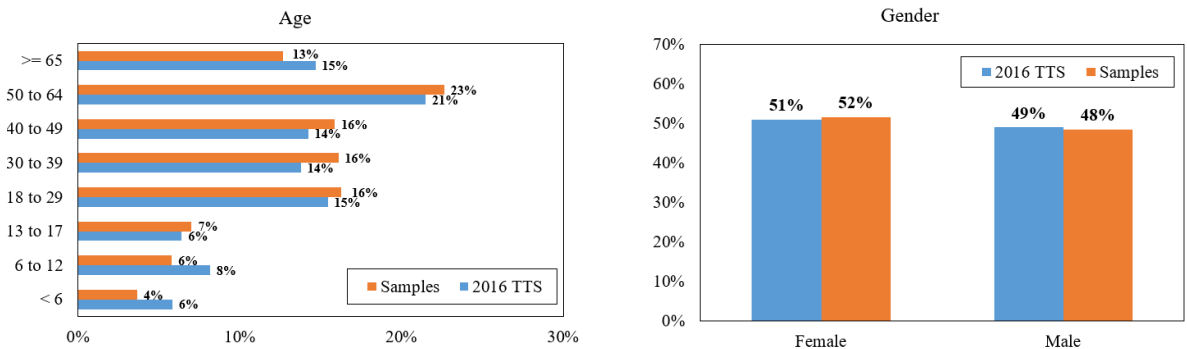

Figure 2. Comparison of key personal attributes – sample vs. 2016 TTS

**Weekly spending on e-shopping**

As part of the 2021 COVHITS survey, respondents were asked to provide information on their use of e-shopping in the week (Monday to Sunday) preceding the survey. Key summary statistics are summarized in Figure 3. Overall, one-third of respondents did not receive any home deliveries in the week before the survey, while 43% received one or two deliveries. On average, respondents received 1.72 home deliveries in the week preceding the survey; among households that received at least one delivery, the average was 2.57 home deliveries per household. This is consistent with the work of Barua et al. (19) and Dias et al. (18), who found that home deliveries are a relatively rare occurrence for most households. Besides, durable items (such as books, clothing, and electronic devices) were the most common item received via home delivery, followed by cooked meals and groceries. In terms of spending on e-shopping, most respondents spent around $100 in the week on e-shopping orders; however, a fair portion of respondents spent more than $200 on e-shopping each week. On average, respondents spent an average of $298.80 through e-shopping services in the week preceding the survey. When zero-delivery households are omitted, the average increases to $444.50. On each delivery, more than 50% of the delivery were worth less than $50.

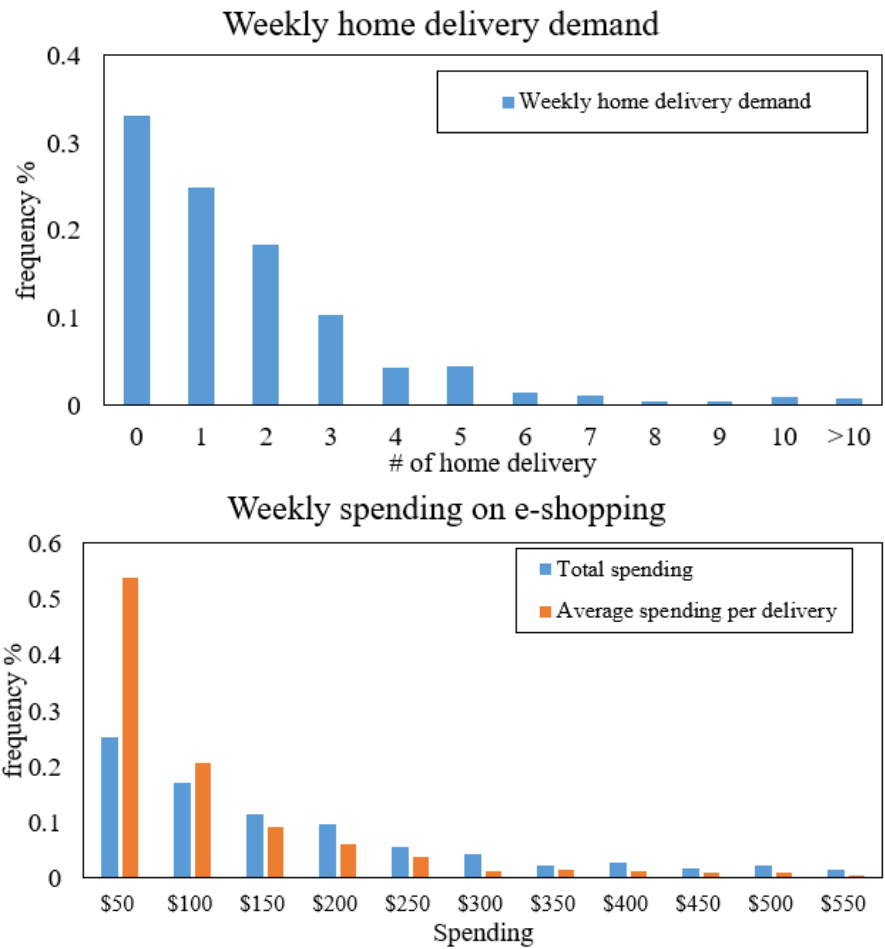

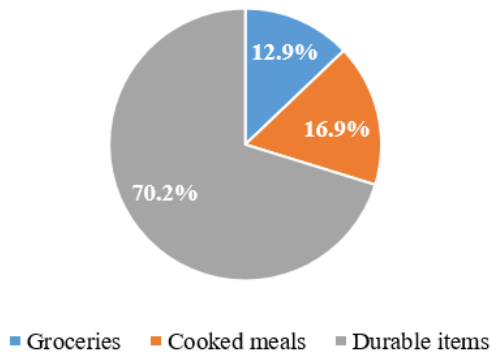

Figure 3. Summary statstics of household weekly e-shopping demand

## 5.0 Modelling Results

In this section, three models (RF, XGBoost & ordered probit model) are established to predict households' weekly home delivery demand. Performances will be compared across all models and discussed in this section. To estimate all models, 70% of the samples were used as a training set, and 30% were held out as a testing set.

## 5.1 Variable Selection for ML Models

Variable selection determines the variables included in the final ML models. An optimal set of input variables can help build more efficient models and reduce the possibility of overfitting (19). This study utilizes recursive feature elimination (RFE) to select the optimal set of variables. The RFE technique is designed to rank the potential variables based on their importance during model development and eliminate the least important variables without sacrificing prediction accuracy. RFE is an iterative approach. It discards one variable at a time, then trains the model with the remaining variables and ranks them by importance. It will discard the next variable until an optimal number of variables remain.

Through RFE, 87 variables are selected from more than 100 potential variables initially prepared for the models. The selected variables can be divided into three categories: socioeconomic variables, trip attributes, and built environment (land use) variables. The built environment variables include the number of Enhanced Points of Interest (EPOI) and land use area of different types of businesses and services summarized for each traffic analysis zone within the study area. Table 1 summarizes the selected variables along with their descriptions.

Table 1. Description of selected variables into ML models

| Variable Name | Description |
|---|---|
| **Socio-economic Variables** | |
| HH average age (log) | Logarithm of average household age |
| HH income (log) | Logarithm of household annual income |
| HH gender | Number of household numbers who are male / female |
| HH male percentage | Percentage of males in the household |
| HH driver | Number and percentage of household members who have a driver's licence |

| HH worker | Number and percentage of workers in the household |
|---|---|
| HH WFH only FT/PT worker | Number of household members who only work from home full-time / part-time |
| HH WFH hybrid FT/PT worker | Number of household members who work full-time / part time and work from home sometimes |
| HH work outside FT/PT worker | Number of household members who work full-time / part-time and never work from home |
| HH retired | Number of household members who are retired and not in labor market |
| HH worker schedule | Number of household members who have a fixed / flexible / output-oriented work schedule |
| HH post secondary student | Number of household members who are post secondary students |
| HH kids | Number of household members who are age under 12 / between 13-17 / under 18 |
| HH kid percentage | Percentage of household members who are age under 18 |
| HH adults | Number of household members who are age between 18-64 / 64+ |
| HH adult percentage | Percentage of household members who are age above 18 |
| HH size | Number of people in the household |
| HH university degree | Number and percentage of household members with a post secondary degree |
| HH high school degree | Number and percentage of household members whose highest degree is high school |
| HH adult manual bikes | Number of adult manual bikes in the household |
| HH vehicles | Number of vehicles in the household |
| HH adult E-bikes & scooters | Number of E-bikes and scooters in the household |
| HH income | Household annual income is between $15k - $39k / $40k - $59k / $60k - $79k / $80k - $99k / $100k - 125k / $150k - 199k / $200k + |
| HH structure | The household structure is living along / couple only / couple with children / living with parents or grandparents / living with roommates |
| HH tenure type | Tenure type of the household is rent / own |
| HH dwelling type | Dwelling type of the household is house / townhouse / apartment |
| HH region | The household lives in Toronto / Peel / York |
| HH car share | 1: if any member of the household has a car share membership |
| HH bike share | 1: if any member of the household has a bike share membership |
| Never shop online | 1: if members of the household never shop online / 0: otherwise |
| Online grocery membership | 1: if the household have memberships for online grocery shopping services / 0: otherwise |
| **Trip Attributes (computed using trip records collected from the last workday)** | |
| HH total travel distance (log) | Logarithm of the total distance (in km) traveled by all household members |
| HH total trip | Total number of trips made by all household members |

| HH shopping trip | Number of shopping trips made by all household members |
| --- | --- |
| **Built Environment & Land use** | |
| EPOI variables (log) | Logarithm of the number of EPOI points (from selected categories) within the traffic analysis zone where the household is located |
| | The following EPOI categories are selected: construction, manufacturing, wholesale trade, retail trade, transportation, information, finance, real estate, professional, administrative & support, education, health care, recreation, hotel & food services, other services, public services, |
| Land use variables (log) | Logarithm of the total area (in m$^2$) belonged to selected land use categories within the traffic analysis zone where the household is located |
| | The following land use categories are selected: commercial, government & institutional, open area, park & recreational, residential, resource & industrial, waterbody, total area |
| Population density | Population density of the traffic analysis zone that the household is located |

## 5.2 Ordered Probit Model

Table 2 shows the ordered probit home delivery frequency model estimated using the training dataset. Multiple specifications were tested, and the final specification yields the best model fit. Most estimated coefficients having the expected signs are statistically significant at the 95% confidence level. The McFadden $R^2$ of 0.11 indicated reasonable goodness-of-fit. The lower Akaike information criterion (AIC) of the full model also indicates better model fit.

On the household level, household tenure, workers' working status, and online grocery memberships will increase the probability of higher home delivery demand. In addition, living in a rented property and having workers with flexible working schedules will lead to higher home delivery demand. On the other hand, living in an apartment, older average ages, and a larger percentage of males in the household will negatively affect higher home delivery demand. These findings conform with literature that millennials, women, individuals from the higher-income household are more likely to shop online (17).

For mobility tools, owning bike-sharing memberships and electric bikes will increase the probability of higher home delivery demands. These findings fit expectations, because the users of shared mobility and electricity-powered micro-mobility tools are most likely to be tech-savvy and in favour of using e-shopping. Ramirez (17) also found that people who commute using active modes are more likely to use e-shopping. At the same time, having a higher percentage of drivers in the household will negatively affect home delivery demand. This is consistent with the literature (13,14,18), indicating substitution effects between home delivery and travelling.

For land-use attributes, more educational and recreational institutions in the zone where the households were located will increase the delivery demand. This finding confirms previous observations regarding household average age. Households located near schools and recreational facilities might be younger households. On the other hand, more public service and commercial institutions in the zone in which the household is located will contribute to lower delivery demand. More specifically, households in the Region of Peel (a subregion in GTA) had the lowest home delivery demand compared to households in other regions. The average weekly home delivery demand in the training dataset was 1.66 per household. However, the average demand in the Region of Peel was only 1.53 per household per week.

A dummy variable capturing whether the household reported never using e-shopping services during the past year was included in the model. This variable was included to distinguish structure zeros from incidental zeros in the dataset. As shown in Figure 3, around 30% of the households did not report any home delivery. However, it is possible that these households just happened to receive zero home delivery during the survey period. In this case, those households should be classified as incidental zero.

On the other hand, structural zero happened when households behaviourally exclude e-shopping from their shopping channel. Therefore, households that did not make an online purchase in the past year should be a reasonable estimator of such cases. The positive and large coefficient of the dummy variable indicates a strong effect of structure zero. However, the coefficient is not significant. The reason can be attributed to the small sample size. Only 27 out of 595 records were structural zeros. Nevertheless, the dummy variable is kept in the model considering its strong effect and the fair possibility of its statistical significance with increasing sample size.

Table 2. Ordered probit weekly home delivery frequency model

| Attributes *(attribute codes in ML model)* | Estimate | t-value |
|---|---|---|
| **Household attributes** | | |
| Tenure as rental *(HH_tenure_rent)* | 0.26 | 2.33 |
| Dwelling types as apartment *(HH_dwelling_types_apt)* | -0.21 | -1.90 |
| Income between $15,000 and $39,999 *(HH_income_15_39)* | -0.29 | -2.29 |
| Percentage of male *(HH_male_precentage)* | -0.41 | -3.12 |
| Log of average age *(HH_average_age_log)* | -0.50 | -3.95 |
| Workers with flexible working schedule *(HH_worker_schedule_flex)* | 0.16 | 2.10 |
| Having online grocery membership *(Online_grocery_membership)* | 0.73 | 6.39 |
| Never shopped online during past 12 months *(Never_shop_online)* | -4.76 | -0.43 |
| **Mobility tools** | | |
| Having bike sharing membership *(HH_bike_share)* | 0.44 | 2.46 |
| Having 2 E-bikes *(HH_E-bikes_scooters)* | 0.53 | 1.51 |
| Percentage of licensed drivers *(HH_driver_percentage)* | -0.25 | -1.71 |
| **Land use attributes** | | |
| Log of educational points of interest *(EPOI_Education_log)* | 0.14 | 2.12 |
| Log of recreational points of interest *(EPOI_Recreation_log)* | 0.09 | 1.30 |
| Log of public services points of interest *(EPOI_Public_log)* | -0.36 | -3.03 |
| Log of area of commercial land *(LU_Commercial_area_log)* | -0.01 | -1.33 |
| Located in Peel Region *(Peel)* | -0.22 | -1.85 |
| 1|2 | -1.71 | -3.50 |

| | | |
|---|---|---|
| 2\|3 | -1.11 | -2.29 |
| 3\|4 | -0.60 | -1.23 |
| 4\|5+ | -0.33 | -0.68 |
| | | |
| Log likelihood (full) | -849.0 | |
| Log likelihood (null) | -954.4 | |
| McFadden's $R^2$ | 0.11 | |
| AIC (full) | 1740.0 | |
| AIC (null) | 1918.8 | |

(1) McFadden's $R^2$ is calculated as 1- ($LL/LL_{null}$), where LL is the log-likelihood and $LL_{null}$ is Log likelihood of null model (2) Akaike Information Criterion (AIC) is calculated as -2(LL-k), where k is the number of estimated parameters including intercept. (3)

## 5.3 Prediction performance

Performances of all models were examined using the testing dataset. The observed and predicted distribution from each model is presented in Figure 4. The results indicate that the econometric model (ordered probit) was able to predict the aggregate distribution in the testing dataset. On the other hand, both ML models tend to over predict the majority class, which corresponds to having zero home delivery demand. The RF model had the worst performance, over-predicting the share of zero-delivery households by around 25%. Imbalanced datasets are a common issue in ML applications (24,25). Unlike econometric model that matches the aggregate distribution in the testing dataset, ML models tend to over predict the majority class in unbalanced datasets to maximize their accuracy of modelling outcomes. However, this approach will jeopardize the model's capability to produce unbiased predictions. Consequently, the aggregate distributions predicted by RF and XGBoost models have deviated from the observed distribution.

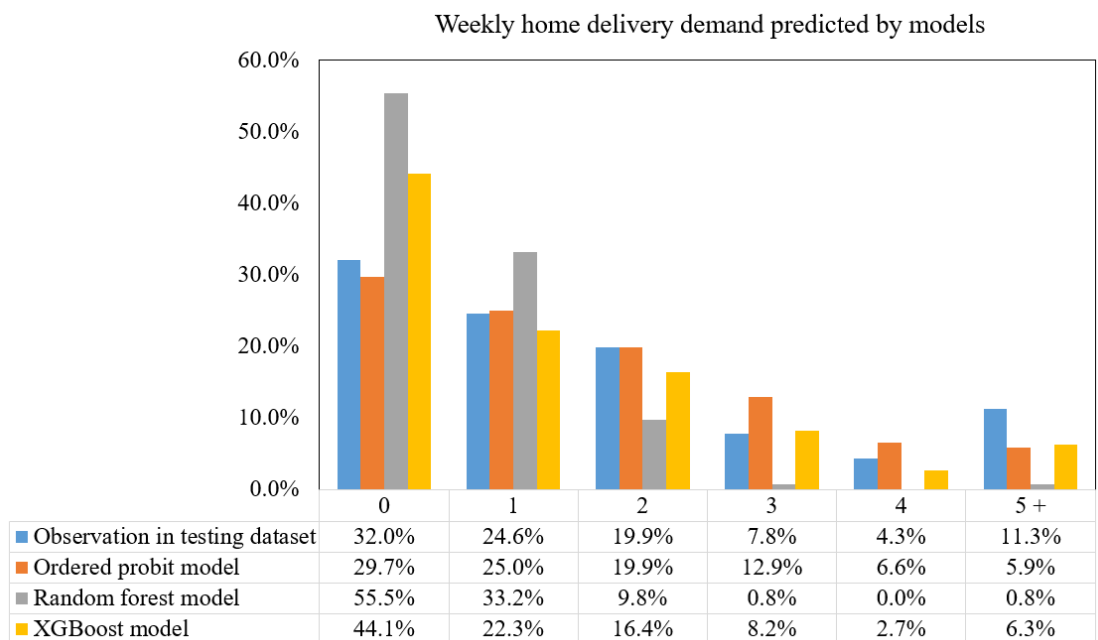

Figure 4. Weekly home delivery demand predicted in testing dataset by models

Besides the aggregate distribution, model performance was also examined at the disaggregate level using confusion matrices. Results are presented in the form of heat maps generated based on recall accuracy in Figure 5. The overall prediction accuracy was 30%, 34%, and 33% for ordered probit, RF, and XGBoost models, respectively. Surprisingly, the RF model scored the highest overall accuracy, despite its relative inability to predict the aggregate distribution. However, the confusion matrix reflects that the RF model achieved such an accuracy rate by heavily overpredicting the zero-delivery class. At the same time, its recall accuracy for the 3, 4, and 5 delivery classes were merely 0%, 0%, and 3.4%, which is extremely low. On the other hand, the ordered probit and XGBoost models performed reasonably well across all classes.

In summary, the ordered probit model had the best performance among the models tested in this study. It predicted the aggregate distribution of household delivery demand very well and achieved similar recall accuracy as the ML models. Also, it is worth noting that the econometric model only contains 16 variables, whereas ML models contain 87 variables. Such simplicity also has cost-efficiency implications because it requires fewer efforts in terms of data collection and preparation.

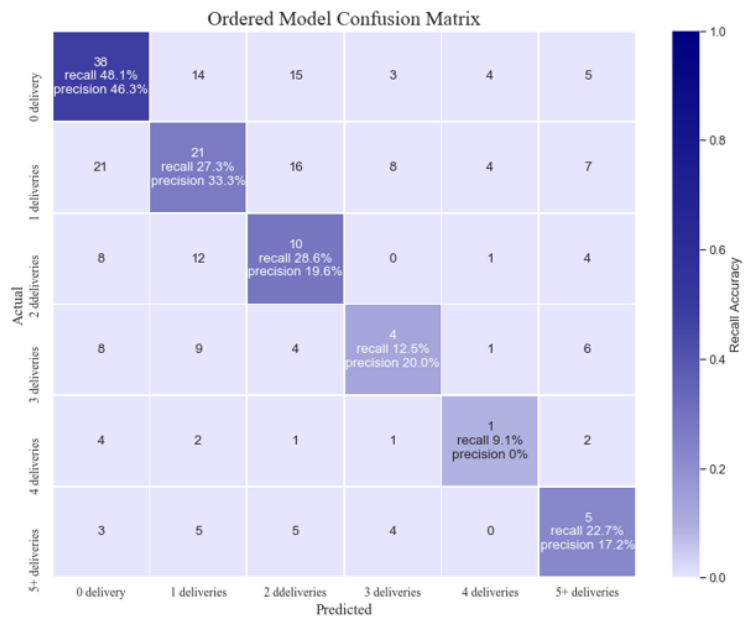

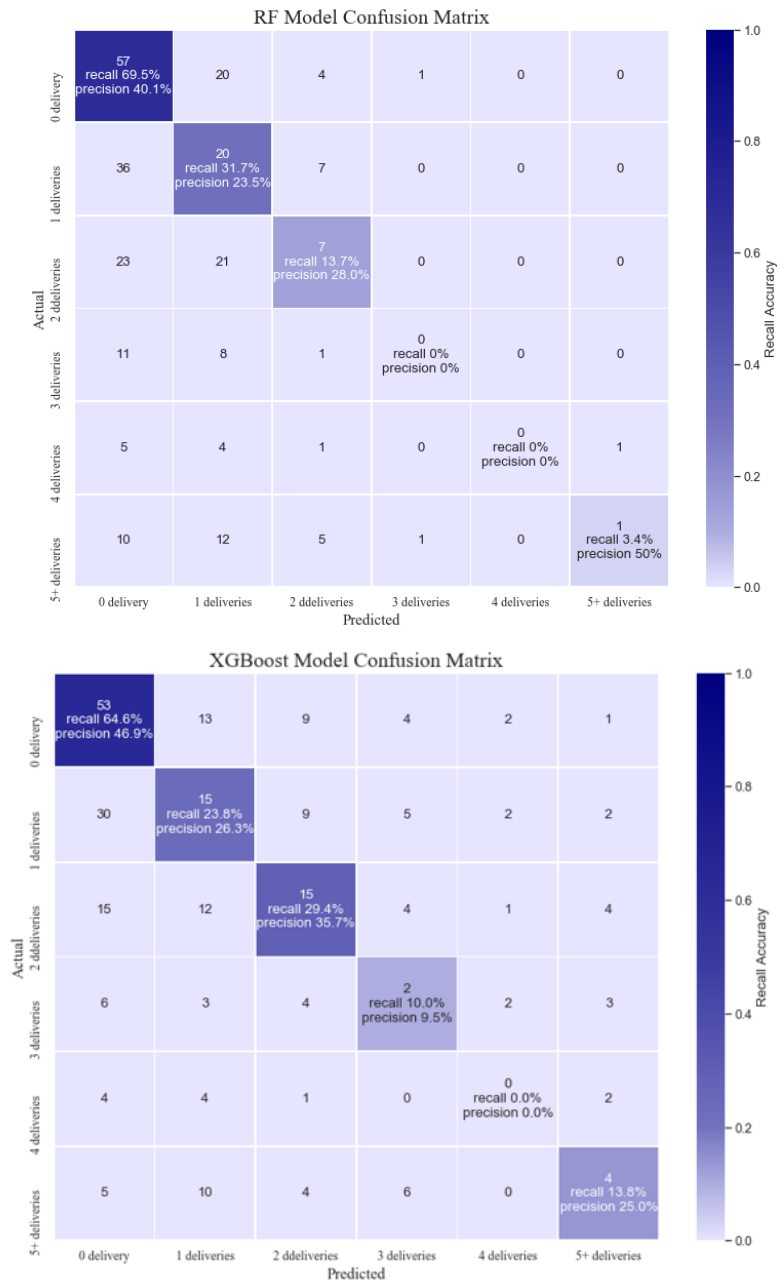

Figure 5. Confusion matrix for weekly home delivery frequency models

## 5.4 Post-modeling Analysis

One of the objectives of this study is to understand the impacts of the variables on the model results. In ML models, the effects of variables can be interpreted from feature importance and feature dependence plots. For the ordered probit model, the effects of variables can be interpreted from coefficient signs and marginal effects. Interpretation results across all modes will be compared and discussed.

**Shapley Additive exPlanations (SHAP)**

This study utilizes the Shapley Additive exPlanations (SHAP) method to compute feature importance and feature dependence. The SHAP value is developed based on the Shapley values from coalitional game theory (26). The Shapley value is defined as the average marginal contribution of a variable. It is measured by the contribution that the variable made to the prediction in each observation compared to the average prediction (27). The detailed formulation of SHAP value is presented below (26,27):

$$g(z') = \emptyset_0 + \sum_{j=1}^{M} \emptyset_j z_j' \tag{8}$$

$$\emptyset_j = \sum_{S \in \{x_1, \ldots, x_p\} \setminus \{x_j\}} \frac{|S|! \, (p - |S| - 1)!}{p!} \left( F\left(S \cup \{x_j\}\right) - F(S) \right) \tag{9}$$

were,
$z'$ = the coalition vector, $z' \in \{0,1\}^M$
$M$ = the maximum size of the coalition vector
$\emptyset_j$ = the Shapley value
$S$ = subsets of the input variables
$x$ = vector of the variables values for the observation of interest
$p$ = the numbe of input variables
$F$ = the trained model

The SHAP value represents the contribution of an input variable to the model result. A larger SHAP value means that the variable plays a more important role in predicting the outcomes.

**ML Model Feature Importance**

Feature importance is used to quantitively examine the contribution of the input variables on the predictions of the model. SHAP feature importance is utilized in this study. Since the SHAP value is computed for each individual observation, the global importance of a variable is computed by summing up the absolute SHAP values across the dataset, as demonstrated by the equation below (27).

$$I_j = \sum_{i=1}^{n} \left| \emptyset_j^i \right| \tag{10}$$

where $i$ represents an observation, and $j$ represents a variable. A variable with a larger absolute SHAP value is ranked higher on the importance plot because it has made a larger contribution to the predictions of the model.

Figure 6 illustrates the feature importance plots for the RF and XGBoost models. The plots only display the top twenty most important variables for each model. The top twenty most important variables for the two models are very similar, although the rankings are slightly different. Figure 6 also shows the change in ranks between the two models.

Results from the ML model feature importance plots match closely with the econometric model. Online grocery membership and household average age are the two most important variables in both ML models. This matches with the results of the ordered probit model, where these two variables are statistically significant. Both variables contribute the most when classifying zero home delivery. The percentage of male household members also plays an important role when classifying the zero-delivery class, and it is also statistically significant in the ordered probit model. Several land-use variables are among the top twenty most important variables in both ML models. It is reasonable that land-use variables have higher importance ranks. Greater accessibility to retail stores or commercial areas may influence households' online shopping frequency. Households that live in the same zone as shopping centers and restaurants may shop more in-store than online, given the convenience to do so. Lastly, the RF model also recognized the effect of structural zeros discussed previously in the ordered probit model.

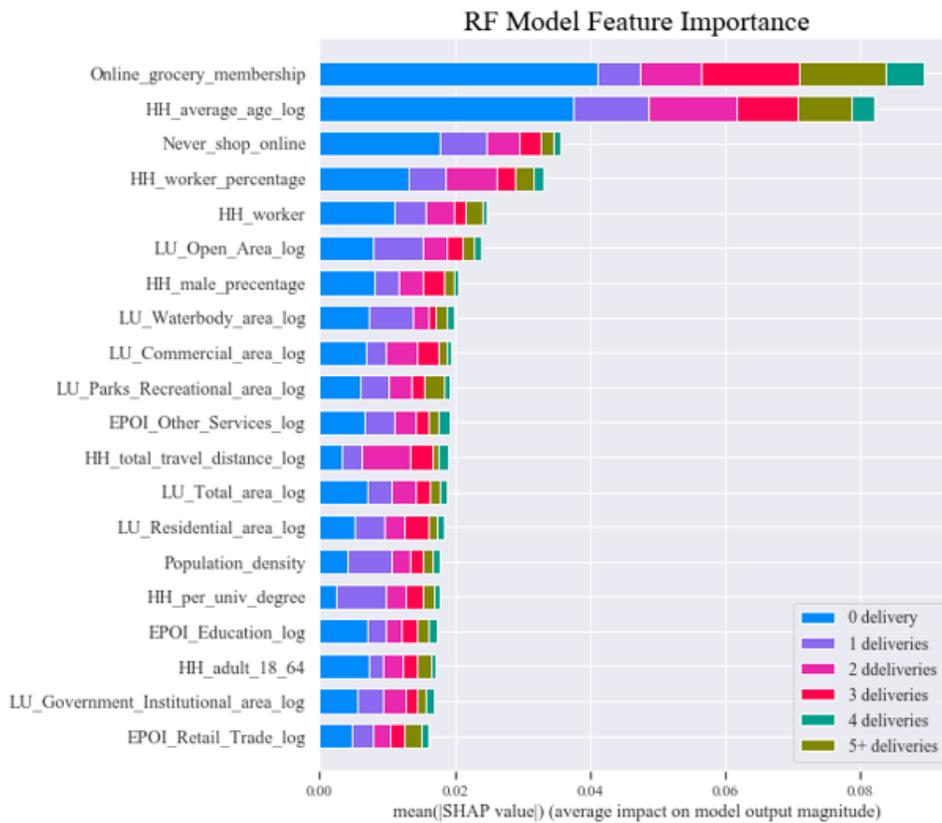

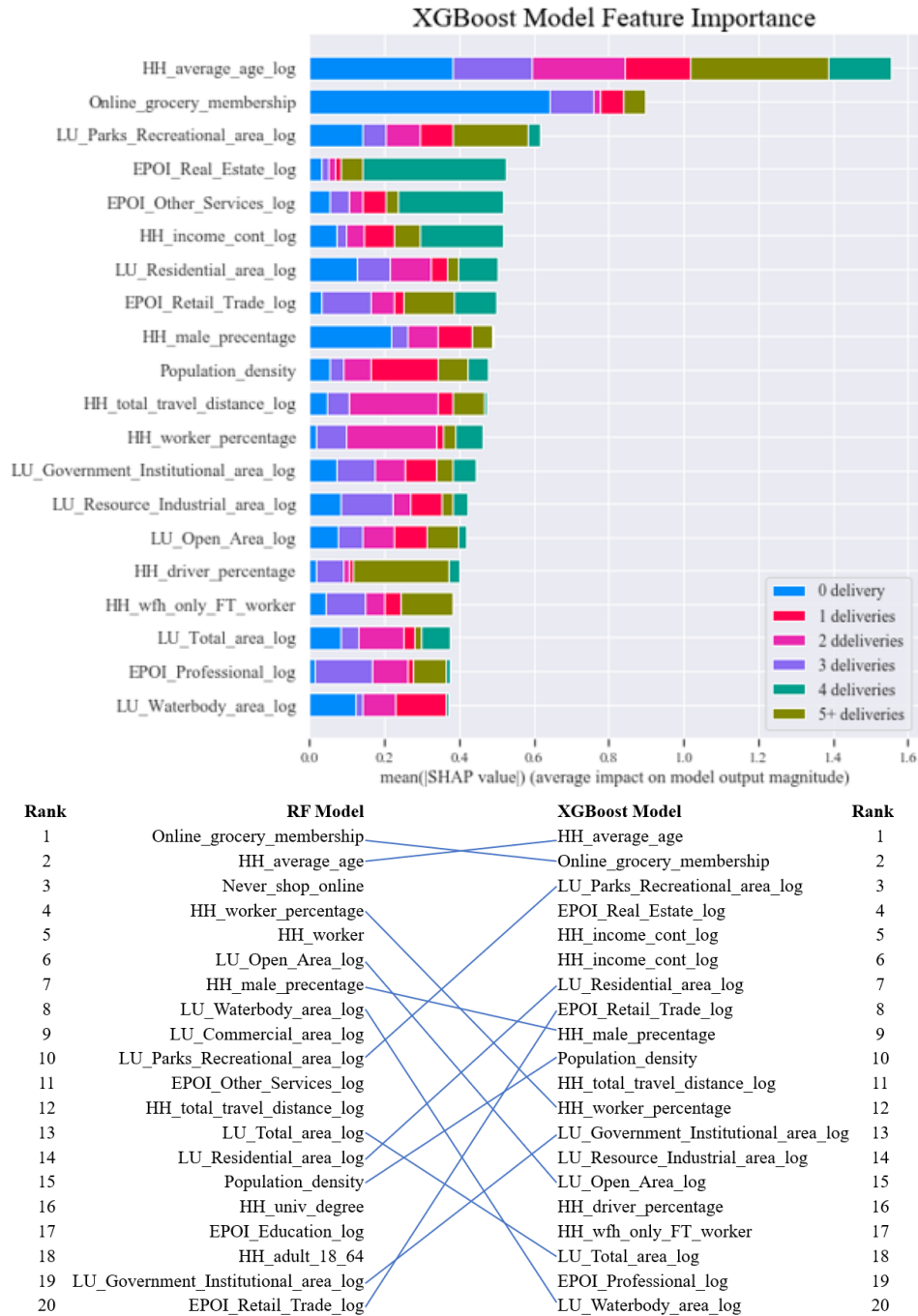

Figure 6. ML model feature importance ranking & comparison

**Marginal Effects & Feature Dependence**

Feature importance only indicates the level of importance for each variable in ML models. However, it does not indicate the direction of effects. This section will compare the effects of selected variables discussed in the previous section between ML and ordered probit models.

The marginal effects are a common measurement used to interpret behavioural effects of variables in econometric models. Direct marginal effects capture the change in probability of observing the dependent variable given one unit change of one independent variable. The equation to calculate direct marginal effects in probit models is:

$$Marginal\ Effect = \frac{\partial Pr(y = j|x)}{\partial x} = \frac{\partial G(\beta x)}{\partial x} = g(\beta x)\beta \tag{11}$$

where $g(z)$ is the probability of standard normal distribution.

In the ordered probit model, marginal effects were calculated to investigate the effect of the average age of household members, having online grocery memberships, and the percentage of males in the households on the households' home delivery frequency (see Figure 7). The online grocery membership variable is a binary indicator reflecting whether the household has online grocery shopping memberships. The average household age and percentage of males are continuous variables but differ in nature. Also, they are all statistically significant in the ordered probit model (see table 2) and have high importance in ML models. Thus, they will serve the purposes of comparing variables effects between the econometric model and ML models well.

The marginal effects reflect that having online grocery memberships (which allows people to shop for groceries online and enjoy free home delivery when certain conditions are met) will decrease the probability of having zero-delivery demand and increase the probability of having more than two deliveries per week. In contrast, an increase in households' average age will improve the likelihood of having zero-delivery demand. Lastly, a higher percentage of males in the household will also lead to a higher probability of having zero home delivery demand.

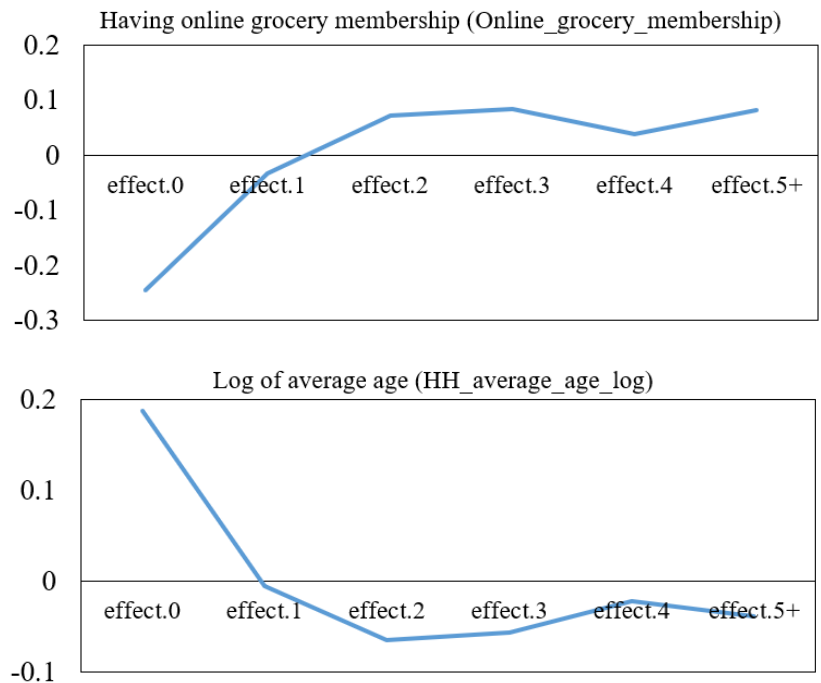

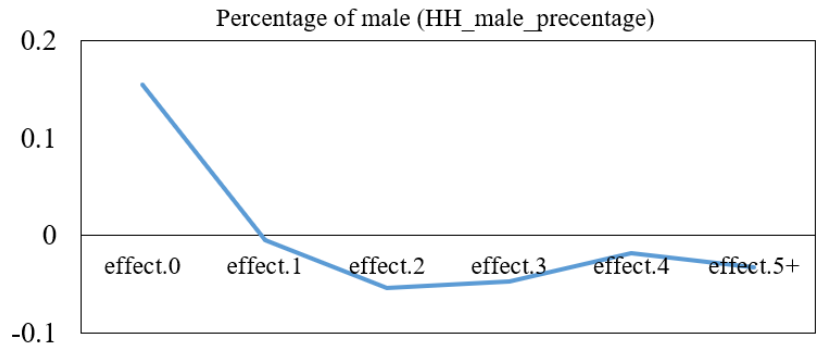

Figure 7. Marginal effects of selected variables in the ordered probit model

For the ML models, feature dependence shows the effects of the variables. This study uses the SHAP dependence plot because it can capture the variance of the effects instead of just presenting the average effects (27). It is generated by plotting the SHAP values of all observations for a given variable against the variable values. For classification problems, a SHAP dependence plot can be plotted for one variable and one class at a time. This study includes six classification categories, so six SHAP dependence plots will be generated per variable.

The marginal effects of the ownership of online grocery memberships, average household age, and percentage of male members in the household are calculated for the ordered probit models. The SHAP dependence plots for the three variables are presented in Figure 8. The results indicate that ML models and the ordered probit models revealed precisely the same variables effects. Moreover, the SHAP dependence plots can provide richer information than the average marginal effects.

From the dependence plots, it can be observed that not having an online grocery shopping membership can increase the probability of zero home delivery demand. At the same time, it can decrease the probability of having more than two deliveries per week. This finding fits the observation from the ordered probit model. Taking a deeper look, the dependence plots generated by the RF model and the XGBoost model for this variable are very similar, except for having four deliveries. The XGBoost model's plot for the four-delivery class shows that the SHAP values for all observations are zero. It means that this variable does not contribute to this prediction result, indicating that the XGBoost model does not utilize this variable when predicting observations with four home deliveries. This can be verified by the feature importance plot of the XGBoost model shown in Figure 6. Although this variable is ranked the second-most important variable overall, it is not important when predicting observations with four deliveries.

In both ML models, the likelihood of having zero home delivery demand decreased monotonically with an increasing average age. This seems reasonable because older people are less familiar with the internet and online shopping and prefer in-store shopping. For the one-delivery class, both ML models show the likelihood of having one home delivery per week peaks at a household average age of 40. In the case of three deliveries, slightly different conclusions can be drawn from RF and XGBoost models. The RF model reveals a monotonically decreasing trend of having three deliveries with increasing age. However, the XGBoost model plot peaks at an average age of 30 before decreasing afterward. Moreover, in the case of five and more deliveries, the RF model reveals another monotonically decreasing trend. But the XGBoost model peaked at the age of 40 and demonstrates a sudden discontinuity afterward. This indicates a clear generation gap in terms of e-shopping. The generation of millennials (born after 1980) is generally more receptive to e-shopping and home delivery than older generations. Households made up of millennial adults are at their peak of consumption. Likewise, they are more familiar with the internet and online shopping compared to the older generations. This confirms the literature that millennials are more likely to shop online than in-store (17).

Lastly, in both ML models, a higher percentage of male household members will lead to a higher likelihood of having zero home delivery demand. This finding again fits the observation in the ordered probit model. However, the effects of the variable on having more than one delivery in ML models are rather ambiguous. For the one-delivery class, plots from both ML models peak when the percentage of male members is around 50%. The plots from both ML models show that households with a lower percentage of male members are more likely to receive two or more home deliveries per week. This again confirms findings from the literature that females tend to have a higher likelihood of shopping online (17).

Overall, the findings from the marginal effects (ordered probit model) and the SHAP dependency plots (ML models) are very similar and are consistent with one another. They all conclude that not subscribing to online grocery shopping memberships, having higher average household age and a higher percentage of male members in the household would increase the probability of zero home-delivery demand.

**Online_grocery_membership**

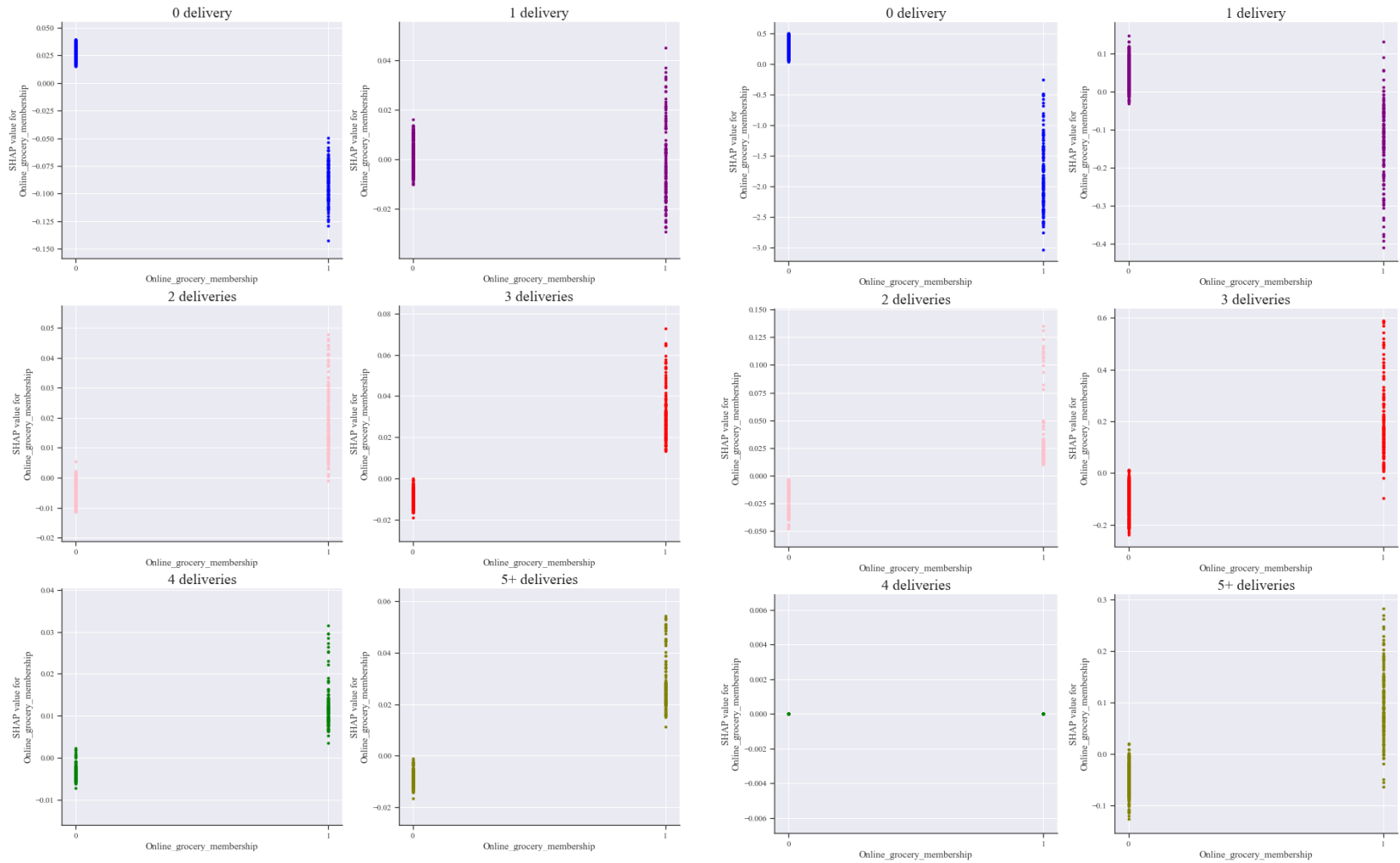

**HH_average_age_log**

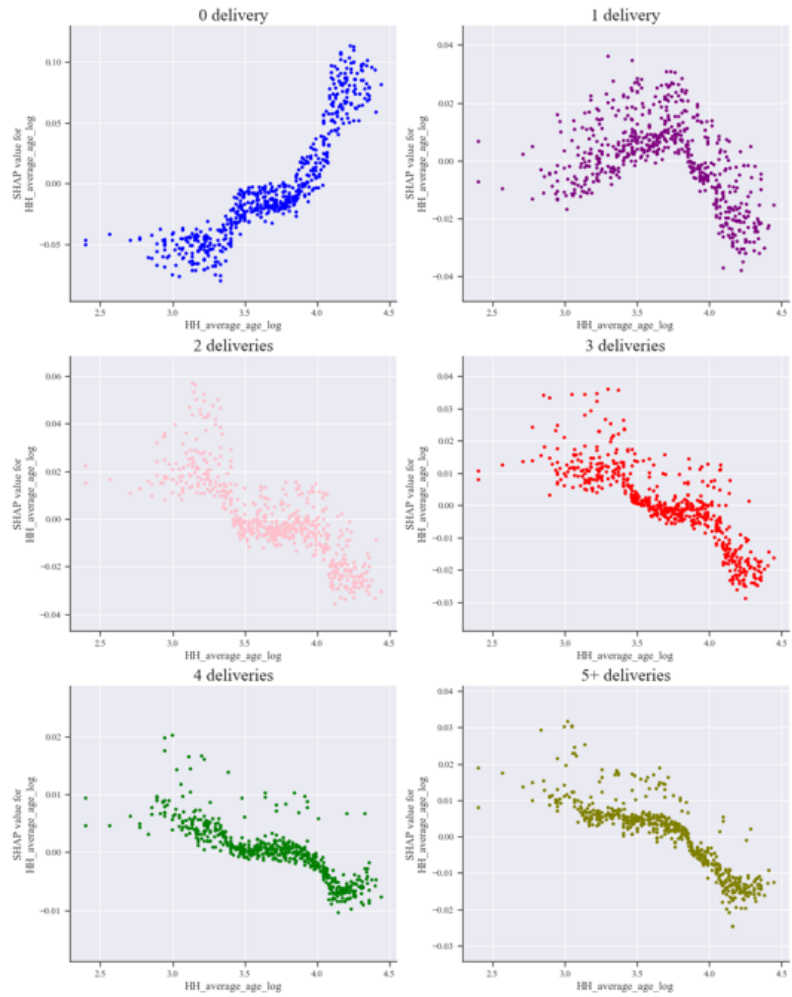
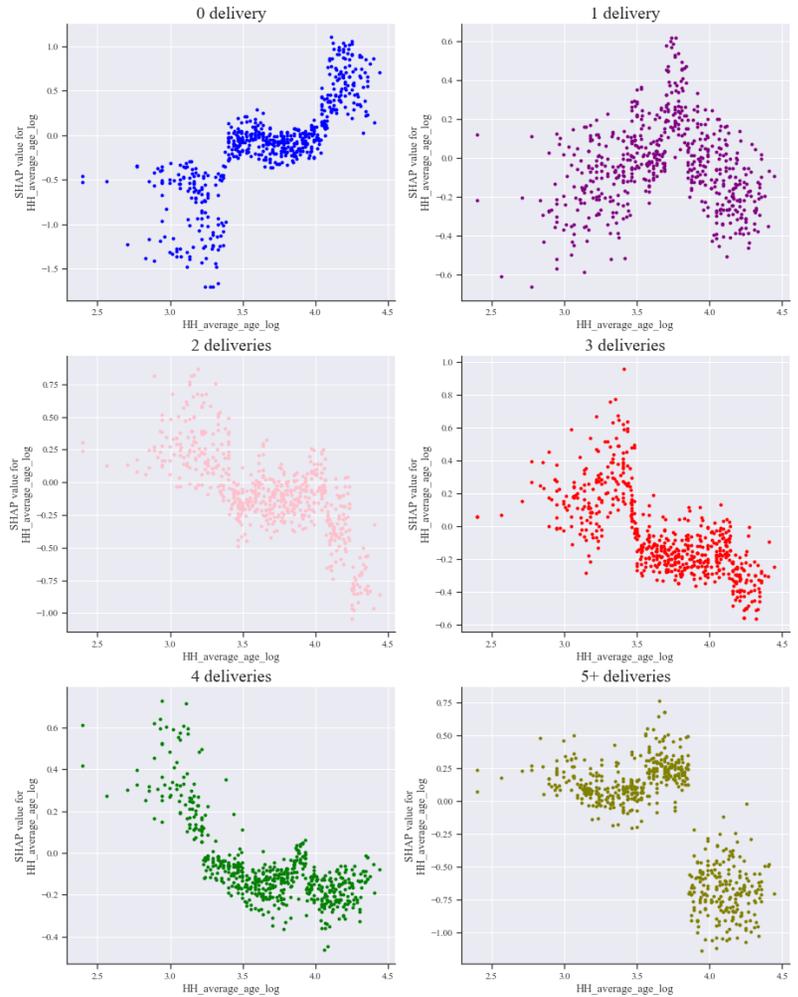

Random forest                                    XGBoost

**HH_male_precentage**

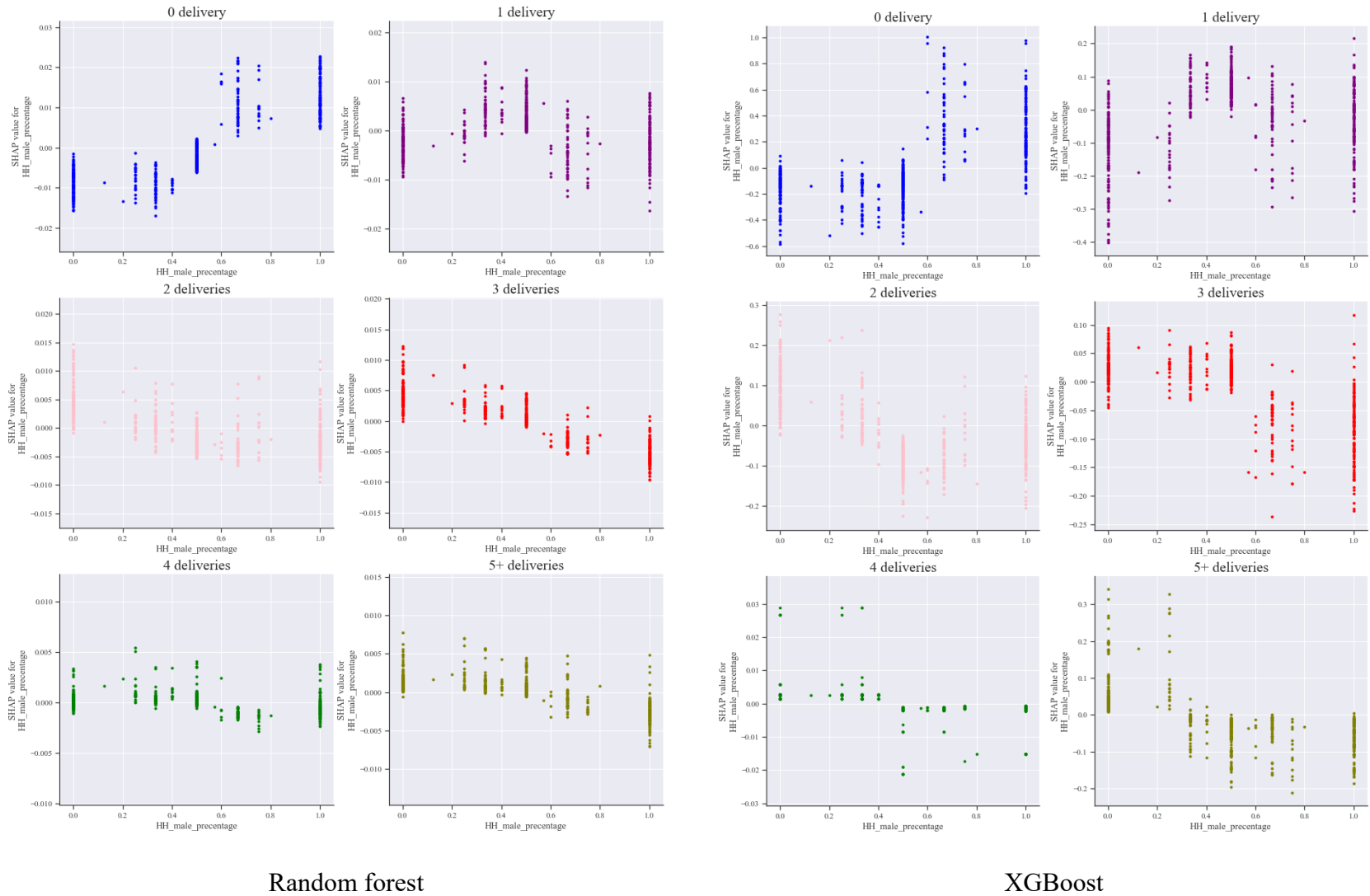

Random forest                                                        XGBoost

Figure 8. Dependence plots of selected variables in ML models

**6.0 Conclusion**

In conclusion, this study developed two ML models (e.g., XGBoost & RF models) and an econometric model (e.g., ordered probit model) to predict weekly home delivery frequencies. In addition, this study found that socioeconomic factors, such as having an online grocery membership, household members' average age, percentage of males in households, and number of workers in the household, and various land-use factors will influence home delivery demand.

This study also provided a fair comparison in model performance between ML models and the econometric model. The study found that the ordered probit model can accurately predict the aggregate distribution of household delivery demand. In contrast, ML models tend to overpredict the majority class (having zero weekly delivery demand in this case). On the disaggregated level, the ordered probit model achieved similar recall accuracy as the ML models. Also, it is worth noting that the econometric model achieved this performance with far fewer variables included in the final model. Such simplicity is advantageous in terms of cost-efficiency, since it requires less data collection and preparation efforts.

This study also examined the interpretability of ML models by comparing their outcomes with the ordered probit model. It found that ML models can provide similar interpretations in terms of variables effects compared to the econometric model. This finding indicated the potential to apply ML models in transport-related applications such as planning and policymaking.

There are limitations associated with this study. The sample size used in this study is relatively small for ML applications. The performance of ML models might improve with a larger dataset. However, many transport-related studies are based on the same sample size as used in this study. Thus, the applicability of ML models with small sample size is an unavoidable issue in transportation planning.

Moreover, the dataset used in this study is rather imbalanced, and the ML models used in this study do not work well with an imbalanced dataset. Techniques to mitigate this issue can be tested in future studies, such as over-sampling the minority classes, under-sampling the majority classes, or double the cost of misclassifying minority classes. Future study can apply advanced econometric model such as two-stage ordered probit model or multinomial logit (MNL) style ordered model. These state-of-the-art econometric models might further improve the performances of econometric models. Nonetheless, the basic ordered probit model presented in this study matched the performances of two state-of-art ML models, which serves the purpose of this study.

**Acknowledgment**

The study was funded by an NSERC Discovery Grant and Percy Edward Hart Professorship Grant. The authors bear the sole responsibility for all results, interpretations, and comments made in the paper.

**Author Contribution Statement**

The authors confirm contribution to the paper as follows: *Study conception and design*: Y. Liu, K. Wang, K.M.N. Habib; *Data collection*: K. Wang; Y. Liu; *Analysis and interpretation of results:* Y. Liu, K. Wang, P. Loa; *Draft manuscript preparation:* Y. Liu, K. Wang, P. Loa;

1  *Overall project supervision:* K.M.N. Habib. All authors reviewed the results and approved the
2  final version of the manuscript.